\documentclass[
    ,final            
  ]
  {aipproc}

\layoutstyle{8x11single}

\begin{document}

\title{The quenched-disordered Ising model in two and four dimensions}


\author{A.~Gordillo-Guerrero}{
  address={Departamento de Ingenieria Electrica, Electronica y Automatica,
 Universidad de Extremadura, Avda Universidad s/n, Caceres, 10071 Spain}
}

\author{R. Kenna}{
  address={Applied Mathematics Research Centre, 
Coventry University, Coventry, CV1 5FB, England}
}

\author{J.J.~Ruiz-Lorenzo}{
  address={Departamento de Fisica,  Universidad de Extremadura,
Avda Elvas s/n, Badajoz, 06071 Spain}
}

\begin{abstract}
We briefly review the Ising model with uncorrelated, 
quenched random-site or random-bond disorder, which  
has been controversial in both two and four dimensions. 
In these dimensions, the leading exponent $\alpha$, 
which characterizes the specific-heat critical behaviour,
vanishes and no Harris prediction for the consequences of quenched disorder can
be made. In the two-dimensional case, the controversy is between the strong
universality hypothesis which maintains that the leading critical exponents 
are the same as in the pure case and the weak universality hypothesis, 
which
favours dilution-dependent leading critical exponents. 
Here the random-site version of the model is subject to a finite-size scaling analysis, paying special attention to the implications
for multiplicative logarithmic corrections. The analysis is fully supportive of the scaling relations for logarithmic corrections and of the strong scaling hypothesis in the 2D case. 
In the four-dimensional case unusual corrections to scaling characterize the model, and the precise nature of these corrections has been debated. 
Progress made in determining the correct 4D scenario is outlined.
\end{abstract}

\maketitle


\section{Introduction}

Continuous phase transitions are characterized by  critical exponents. 
In terms of the reduced temperature $t$ and the reduced external magnetic 
field $h$, which are measures of how far the system is from its critical 
point $(t,h)=(0,0)$, the standard power-law leading behaviour is as follows:
\begin{eqnarray} 
{\mbox{specific heat at $h=0$:}} \quad \quad\quad\quad\quad\quad \quad \quad \quad ~~~ c_\infty(t)    & \sim & |t|^{-\alpha} \,,
 \label{alpha}
\\
{\mbox{spontaneous magnetization at $h=0$, $t<0$:}} \quad m_\infty(t)    & \sim & |t|^{\beta} \,,
 \label{beta}
\\
{\mbox{susceptibility  at $h=0$:}}  \quad \quad\quad\quad\quad\quad \quad \quad  ~~~ ~~ \chi_\infty(t) & \sim & |t|^{-\gamma}  \,,
 \label{gamma} 
\\
{\mbox{magnetization  at $t=0$:}}  \quad\quad \quad\quad\quad\quad\quad \quad  ~~~ m_\infty(h)    & \sim & |h|^{\frac{1}{\delta}} \,,
 \label{delta}
\\
{\mbox{correlation length  at $h=0$:}}  \quad \quad \quad\quad\quad\quad
\quad ~ \, \, \, \xi_\infty(t) & \sim & |t|^{-\nu} \,.
\label{nu}
\end{eqnarray}
The subscripts here represent the linear extent of the system.
The correlation function at criticality decays as 
\begin{equation}
{\cal{G}}_\infty(x)     =  x^{-(d-2+\eta)}       \,,
\label{eta}
\end{equation}
where $x$ represents distance along the lattice,
the dimensionality of which is $d$.
These six critical exponents are related by four scaling relations, namely
 \begin{eqnarray}
      \nu d              & = & 2 - \alpha \, , \label{SR1} \\
      2\beta + \gamma    & = & 2 - \alpha \, , \label{SR2}\\
      \beta (\delta - 1) & = &   \gamma   \, , \label{SR3}\\
      \nu ( 2 - \eta)    & = &  \gamma \,.     \label{SR4}
 \end{eqnarray}

If a pure system (i.e., a system defined on a 
regular intact lattice) is characterized by a particular
set of critical exponents, it is interesting to ask 
what happens to this set if  randomization is 
introduced to the lattice sites or bonds. The quenched
random removal of sites
or the randomization of bond strengths
is believed to immitate the presence of impurities in real physical systems, 
so this question is 
relevant for meaningful comparison with experiments.

In most cases the Harris criterion provides an answer to this question~\cite{Ha74}. If $\alpha > 0$ in the pure system, 
quenched disorder is relevant and the critical exponents  change as 
such disorder is added. If, on the other hand, $\alpha < 0$ in the
pure model, then this type of disorder does not alter critical behaviour 
and the critical exponents are unchanged upon randomizing the 
lattice structure. 
The Harris criterion does not, however,  provide aa clear an answer in the 
circumstance where specific-heat  critical exponent $\alpha$ vanishes
in the pure model, as is the case for the Ising model in both two and four
dimensions. 
As a result, these circumstances have been quite controversial.
In the following two sections the histories and natures of 
the controversies in 2D and 4D are
outlined. 
An approach to tackle these subtle issues is presented in 
the following section where the results of applications of this 
method to both cases are also 
presented. Conclusions are drawn in the final section.

\section{The two-dimensional case}

In the same manner as the pure Ising model in two dimensions
can be formulated as a lattice theory of free fermions,
the randomized model can be formulated as a theory of interacting fermions. 
This interaction can be considered perturbatively about the solvable 
pure case. In  \cite{DD}  Dotsenko and Dotsenko used a truncated form of 
Grassmann field theory together with the replica trick, 
to link the problem to the $N=0$ Gross-Neveu model. 
They derived the following behaviour for the specific heat in the RBIM:
\begin{equation}
 c_\infty(t) \sim \ln{|\ln{|t|}|} 
\,.
\label{loglog}
\end{equation}
They also derived an expression for the scaling behaviour of the 
susceptibility, namely $\chi_\infty(t) \sim t^{-2} 
\exp{(-c|\ln{|\ln{|t|}|}|^2)}$, where $c$ is a constant.
Their value $\gamma = 2$ there was different to that in the pure theory 
(which has $\gamma = 7/4$) and
the notion that the leading critical exponents may change when going from
the pure to the random 2D Ising model became known as the 
{\emph{weak universality hypothesis}\/}.
In fact, the notion of weak universality as advanced by Suzuki
\cite{Su74} is that some exponents may change with dilution, but 
combinations which appear in terms of the correlation length 
(e.g., in $\chi_\infty(t) \sim \xi_\infty^{\gamma/\nu}(t)$)
are dilution independent. I.e., $\beta/\nu$and $\gamma/\nu$
are unchanged (as are $\delta$ and $\eta$).

In \cite{GJ83}, Jug derived two-loop renormalization-group (RG) results
in the  2D (and 3D) RSIM. (For a recent review on the 3D RSIM,
 see \cite{FoHo03}.) He also worked out the exact RG and 
$\epsilon$-expansion along  the  curve in $(n,d)$-space where 
the pure system's specific-heat $\alpha$-exponent vanishes.
Here $n$ is the number of components of the order parameter in $O(n)$
models. This includes the $n=1$ RSIM case  in $d=2$ (as well as $d=4$,
which we shall discuss in the next section).
For the 2D case, he showed\footnote{In a contribution to the 1983 
Geilo School, 
{\emph{``Multicritical phenomena''}},
Jug  presented the work of \cite{GJ83} and
announced that for the 2D RBIM treated
with the Grassmann field theory he had reached
$c_\infty(t) \sim A \ln{|\ln{|t|}|} + B |\ln{|t|}|^{-1}$
with $A$ and $B$ undetermined constants \cite{GJannouncement}.}
that the critical behaviour
is controlled by the pure-model fixed point, so that the leading 
exponents are the same as in the pure case.
In particular, and contrary to the susceptibility results of \cite{DD}, 
Jug's approach gave $\gamma = 7/4$ for the diluted model -- which is the 
same value as in the pure case. 
The notion that the leading critical exponents are unchanged by 
randomizing the lattice structure
 became known as the {\emph{strong universality hypothesis}}. 

For $d=2$, Jug also derived  a change in the logarithmic form of the
heat capacity, either to $|\ln|t||^{\hat{\alpha}}$ with 
$\hat{\alpha}={\cal{O}}(\epsilon)$ or $\ln{|\ln{|t|}|}$ if $\hat{\alpha}=0$
\cite{GJannouncement}.
In \cite{GJ84}, Jug  
confirmed the result (\ref{loglog}) for the heat capacity for the 2D RBIM
using Grassmann field theory.
On the basis of the chronology outlined above, 
we refer to the (now famous) proposed  behaviour  (\ref{loglog})
in the specific heat  as the 
Dotsenko-Dotsenko-Jug (DDJ) double logarithm.

Shalaev later introduced bosonization to the above approaches and 
derived that all critical exponents are the same as in the pure case 
but there are, in fact, non-trivial 
multiplicative logarithmic corrections to the susceptibility and 
correlation length in the RBIM \cite{Boris2D}. 
He derived the exponents of these logarithms and again obtained
the double logarithm of the specific heat, 
results which were later re-obtained by Shankar and Ludwig  \cite{SL}.
Using transfer matrix techniques together with the self-duality of the
RBIM, and the replica trick, to map the model to the $O(N)$ 
Gross-Neveu model, to which RG is applied to one loop, 
Jug and Shalaev later derived 
the logarithmic corrections for the remaining quantities in the RBIM 
\cite{JuSh96}.

Modifying the standard scaling expressions (\ref{alpha})--(\ref{eta}), above,
to include multiplicative logarithmic corrections, we write
the universal scaling forms
\begin{eqnarray}
  c_\infty (t) & \sim & |t|^{-\alpha} |\ln{|t|}|^{\hat{\alpha}}
\,,
\label{c}
\\
  m_\infty (t) & \sim & |t|^{\beta} |\ln{|t|}|^{\hat{\beta}}
\,,
\label{m}
\\
\chi_\infty (t) & \sim & |t|^{-\gamma} |\ln{|t|}|^{\hat{\gamma}}
\,,
\label{chi}
\\
  m_\infty (h)  & \sim & h^{\frac{1}{\delta}} |\ln{|h|}|^{\hat{\delta}}
\,.
\label{mh}
\\
\xi_\infty (t) & \sim & |t|^{-\nu} |\ln{|t|}|^{\hat{\nu}}
\,,
\label{xi}
\\
{\cal{G}}_\infty(x) & \sim &     x^{-(d-2+\eta)}   (\ln{x})^{\hat{\eta}}  
\,.
\label{corrfun}
\end{eqnarray}
Allowing also for the possibility of logarithmic corrections to the 
FSS behaviour of the correlation length, we also write
the universal scaling form
\begin{equation}
 \xi_L(0) \sim L (\ln{L})^{\hat{q}}\,.
\label{qhat}
\end{equation}
A set of universal scaling relations 
for the correction exponents has recently been developed 
which connects the universal hatted exponents in a manner analogous to 
(\ref{SR1})--(\ref{SR4}). These  are \cite{KeJo06a,KeJo06b}
\begin{eqnarray}
  \hat{\alpha} & = & \left\{{\begin{array}{l}
                        ~ 1 + d \hat{q} -  d \hat{\nu}
  \quad  {\mbox{if}} \quad \alpha = 0 \quad {\mbox{and}} 
                                 \quad \phi \ne \pi/4 \\
                        ~ d \hat{q} -  d \hat{\nu}\,  
  \quad  {\mbox{otherwise,}}  
                     \end{array}}\right.
\label{SRlog1}  \\
   2 \hat{\beta} - \hat{\gamma}  & = &  d \hat{q} -  d \hat{\nu} \, , \label{SRlog2}  \\
  \hat{\beta} (\delta - 1) & = &   \delta \hat{\delta} - \hat{\gamma} \, , \label{SRlog3} \\
  \hat{\eta}  & = &  \hat{\gamma} - \hat{\nu} (2 - \eta ) \,. \label{SRlog4} 
\end{eqnarray}
In the first of these, $\phi$ refers to the angle at which the
complex-temperature zeros impact onto the real axis. 
If $\alpha = 0$, and if this impact angle
is any value other than $\pi/4$, an extra logarithm arises in the 
specific heat. This is expected to happen in $d=2$ dimensions,
but not in $d=4$, where $\phi = \pi/4$ \cite{KeJo06b}. 
In \cite{KeJo06b} it was also shown that  
\begin{equation}
\mbox{if} \quad \alpha=0 \quad
\mbox{and if} \quad d(\hat{\nu}-\hat{q})=1 \,,
\label{mloglog}
\end{equation}
then the specific heat necessarily has the double-logarithmic divergence 
(\ref{loglog}).

The Jug-Shalaev-Shankar-Ludwig (JSSL) values for the leading 
critical exponents and for their logarithmic counterparts are
\begin{equation}
 \alpha = 0\,, \quad 
 \beta = \frac{1}{8}\,,\quad 
 \gamma = \frac{7}{4}\,,\quad 
 \delta = 15\,,\quad 
 \nu = 1\,,\quad 
 \eta = \frac{1}{4}\,,
\label{leading}
\end{equation}
\begin{equation}
 \hat{\alpha} = 0\,, \quad 
 \hat{\beta} = -\frac{1}{16}\,,\quad 
 \hat{\gamma} = \frac{7}{8}\,,\quad 
 \hat{\delta} = 0\,,\quad 
 \hat{\nu} = \frac{1}{2}\,,\quad 
 \hat{\eta} = 0\,.
\label{JSSLscaling}
\end{equation}
The observation that the set of leading exponents (\ref{leading}) 
obeys the usual scaling relations (\ref{SR1})--(\ref{SR4})
is a trivial one, since they are identical to those of the pure Ising 
model in $d=2$.
More interesting is the observation that,
with $\hat{q}=0$, \cite{de95,de97Lede06,Aade97StAa97,Aade96},
  the JSSL correction exponents
(\ref{JSSLscaling}) obey the scaling relations for logarithmic corrections 
(\ref{SRlog1})--(\ref{SRlog4}). The observation that they also obey 
(\ref{mloglog}) with $\hat{q}=0$,
leads to a new route to the derivation of the
DDJ double logarithm in the specific heat
\cite{KeJo06b}.
 
Analytically and numerically based alternatives to the JSSL scenario
and to the DDJ double logarithm in the specific heat
have been made in the literature.
Timonin commented that the replica trick, which was employed in the
previous analytical approaches, encounters a problem in that RG is only
valid with 1 or 2 replicas, preventing the necessary $n \rightarrow 0$ limit
\cite{Ti89}.
Using Grassmannian methods and perturbative RG (with certain other 
assumptions but without replicas) he derived $\hat{\alpha} = -1/2$, 
i.e., a {\emph{finite}} specific heat, in the RBIM.
Ziegler used a supersymmetric formulation, where $n=0$ is replaced by 
$N$ bosons and  $2N$ fermions to give a non-perturbative approach  
to show that the specific heat does not, in fact, diverge in the  
RSIM \cite{Zi88} and RBIM \cite{Zi91}. 
Besides the work of Jug, this was the only analytic work on the RSIM
to this point.
Finally, in 1998, Plechko \cite{Pl98} used Grassmann lattice theory to 
give theoretical support for the double logarithm in the RSIM.

To summarize, both the RSIM and the RBIM have been targeted over 
many years
using a plethora of analytical approaches, some of which support the 
weak universality hypothesis and others of which support strong 
universality. 
In order to discriminate between the two, and to 
decide whether or not the specific heat diverges in these models,
there have also been many numerical investigations of the problem.

Early numerical work by Zobin was supportive of the strong hypothesis
in the RBIM in that the susceptibility exponent $\gamma$
was found to be unchanged there \cite{Zo78}.
In 1990, Andreichenko, Dotsenko, Selke and Wang
\cite{AnDo90WaSe90TaSh94} used such a numerical approach
to present evidence in support of the strong hypothesis. 
They focused on the RBIM because the self-duality of that model leads
to an exact value for the critical temperature $T_c$, ameliorating
some aspects of the numerical analyses.  
Since then many numericists claim support for the strong hypothesis 
and the double logarithm in the specific heat, mostly for the 
RBIM, but also for the RSIM. However many others 
support the weak hypothesis 
and a finite specific heat (mostly for the RSIM, but also for the RBIM).
The situation is summarized in Table~\ref{tab1}.
\begin{table}
\begin{tabular}{ll|l|l}  \hline
          &                                                                      & {\bf{ RBIM }}                                                                             & {\bf{ RSIM }} \\
\hline
\multicolumn{2}{l|}{Support strong universality hypothesis }      &  \cite{de97Lede06,Aade96,Zo78,RoAd98,BeCh04,SzIg99LaIg00}   & \cite{Se94,deSt94,ShVa01}  \\
 & $\quad$and theoretical support for $\alpha = \hat{\alpha}=0$ &  \cite{DD,GJ84,Boris2D,SL,JuSh96,KeJo06b}                                          &  \cite{GJ83,GJannouncement,KeJo06b,Pl98} \\
 & $\quad$or numerical  support for $\alpha = \hat{\alpha}=0$     &   
\cite{Aade97StAa97,AnDo90WaSe90TaSh94,HaTo08,WiDo95,FyMa08}&  \cite{HaTo08,BaFe97,SeSh98} \\
\hline
\multicolumn{2}{l|}{Support for weak universality hypothesis }              &   \cite{Ki95}                            & \cite{FaHo92} \\
 & $\quad$and theoretical support for finite $C_\infty(t)$                         &  \cite{Zi91,finiteCinBond}       &  \cite{Zi88} \\
 & $\quad$or numerical  support  for finite $C_\infty(t)$                          &  \cite{Ki00b,Ki00}                            &\cite{He92,KiPa94,Ku94KuMa00,HaMa08} \\
\hline 
\end{tabular}
\caption{Recent works supportive of  the weak or strong scaling hypothesis
and for and against the double logarithm in the specific heat in the 
RBIM and RSIM in $d=2$ dimensions. 
}  
\label{tab1}
\end{table}

While Roder et al. presented strong numerical evidence that 
$\hat{\gamma}=7/8$ in the RBIM \cite{RoAd98}, their series-expansion approach 
did not lead to a clear result for the specific heat, and almost
any reasonable value $\hat{\alpha} \le 1$ could be supported
from their data.
Indeed, it was emphasised in \cite{Ki00b} that  plots 
of the type  contained in Refs.~\cite{AnDo90WaSe90TaSh94,HaTo08} 
for the RBIM
and  in Refs.~\cite{HaTo08,BaFe97,SeSh98} for the RSIM, which 
purport to display double-logarithmic behaviour
of the  specific heat  do not actually  imply 
such divergence. I.e., it is very difficult (even impossible)
to distinguish between 
\begin{equation}
 c_\infty(t) \sim A + B \ln{|\ln{|t|}|}  \quad {\mbox{and}} \quad
 c_\infty(t) \sim A - B |t|^{-\alpha}  \quad {\mbox{or}} \quad
 c_\infty(t) \sim A - B |\ln{|t|}|^{\hat{\alpha}}\,,
\label{controversy}
\end{equation}
with $\alpha < 0$ or $\hat{\alpha}<0$,
on the basis of direct numerical simulations of the specific heat. 
Numerically based counter-claims for the latter two behaviours
(so that the specific heat 
remains finite) 
in the random-bond \cite{Ki00b,Ki00}    and random-site models \cite{He92,KiPa94,Ku94KuMa00,HaMa08} also exist.

Thus, like the analytical situation, the numerical approaches
to the equilibrium 
lattice-disordered Ising models in 2D have generated much 
controversy and debate. 
(For non-equilibrium scaling aspects in these types of models,
see \cite{HePl08}.)
While it is perhaps
fair to say that the strong hypothesis is mostly favoured, 
agreement is not
universal. Also, the double logarithmic behaviour of the specific heat
has resisted attempts at verification because of the difficulties in
disentangling the scenarios of Eq.(\ref{controversy}).

Here we present a method which circumvents these difficulties.
We use the recently developed 
scaling relations for logarithmic corrections \cite{KeJo06a,KeJo06b} to 
express the scaling of Lee-Yang zeros (in particular their
density) in terms of the specific
heat exponents $\alpha$ and $\hat{\alpha}$. Since the density of zeros
is not accompanied by a constant or homogeneous term, it provides
a cleaner Ansatz with which to extract $\alpha$ and $\hat{\alpha}$
numerically, at least in the $d=2$ case.

\section{The four-dimensional case}

Since the upper critical dimensionality of the RSIM, like its pure
counterpart, is $d=4$, the leading critical exponents are given by 
mean field theory
\begin{equation}
 \alpha = 0\,, \quad \beta = \frac{1}{2}\,, \quad \gamma = 1\,, \quad \delta = 3\,, \quad \nu = \frac{1}{2}\,, \quad \eta = 0\,, \quad \Delta = \frac{3}{2}\,,
\label{MF}
\end{equation}
and there is no weak universality hypothesis.
However, this model is characterised by unusual corrections to scaling
together with multiplicative logarithmic terms,
the precise nature of which are unsettled.

The consensus in the literature is that  the scaling behaviour of the
 RSIM in four dimensions is given by
\cite{GJ83,Ah76,Boris4D,GeDe93,BaFe98,HeJa06}
\begin{eqnarray} 
C_\infty(t)    & \approx & A - B|t|^{-\alpha}    \exp{ \left( {-2  \sqrt{\frac{6}{53}|\ln{|t|}|} }\right) }     |\ln{|t|}|^{\hat{\alpha}} \,, 
 \label{C1}
\\
\chi_\infty(t) & \sim & |t|^{-\gamma}          \exp{\left({\sqrt{\frac{6}{53}|\ln{|t|}|} }\right)}         |\ln{|t|}|^{\hat{\gamma}} \,, 
 \label{chiinfty} 
\\
\xi_\infty(t) & \sim & |t|^{-\nu}          \exp{\left(\frac{1}{2}{\sqrt{\frac{6}{53}|\ln{|t|}|} }\right)}          |\ln{|t|}|^{\hat{\nu}} \,.
\label{xiinfty}
\end{eqnarray}
Modification of the theory presented in \cite{KeJo06a} gives
that the scaling behaviour for the magnetization in this 4D model is
\begin{eqnarray}
m_\infty(t)   & = & t^{\beta}        \exp{\left({-\frac{1}{2}\sqrt{\frac{6}{53}|\ln{|t|}|} }\right)}       |\ln{t}|^{\hat{\beta}} \,, 
\label{m1}
\\
m_\infty(h) & = & h^{\frac{1}{\delta}}  |\ln{h}|^{\hat{\delta}} \,.
\label{m2}
\end{eqnarray}
There is no dispute in the literature regarding
 the unusual exponential 
correction terms in (\ref{C1})--(\ref{xiinfty}), but there are
 {\emph{five}} different sets of 
predictions for the exponents of the logarithmic terms,
which differ from their counterparts in the pure model.

\begin{table}
\begin{tabular}{llllllllll}
\hline
    \tablehead{1}{l}{b}{Log exponent}
  & \tablehead{1}{l}{b}{$\hat{\alpha}$}
  & \tablehead{1}{l}{b}{$\hat{\beta}$}
  & \tablehead{1}{l}{b}{$\hat{\gamma}$}
  & \tablehead{1}{l}{b}{$\hat{\delta}$}
  & \tablehead{1}{l}{b}{$\hat{\nu}$}
  & \tablehead{1}{l}{b}{$\hat{\eta}$}
  & \tablehead{1}{l}{b}{$\hat{q}$} 
  & \tablehead{1}{l}{b}{$\hat{\Delta}$}  \\
\hline
Aharony \cite{Ah76}   & {\bf{0.5}}  & 0.25         & ~{\bf{0}}     & 
                0.167 & ~{\bf{0}}    & 0            & 0.125        & 0.25 \\
Shalaev \cite{Boris2D}& {\bf{1.237}}& 0.434        & {\bf{-0.368}}& 
                0.167 & -0.189      & {\bf{0.009}} & 0.120        & 0.803 \\
Jug \cite{GJ83}        & {\bf{0.5}}  & 0.252        & ~{\bf{0.005}} &
                0.170 &             &              &              & 0.248\\
Geldart \& De'Bell \cite{GeDe93} 
                      & {\bf{1.246}}& 0.439        & {\bf{-0.368}}& 
                0.170 & -0.187      &  {\bf{0.005}}& 0.125        & 0.807\\
Ballesteros et al \cite{BaFe97} 
                      & {\bf{0.5}}  & 0.255        & ~{\bf{0.009}} & 0.173
                      & ~{\bf{0}}    & 0.009        & {\bf{0.125}} & 0.245\\
\hline
\end{tabular}
\caption{Analytic predictions for the logarithmic correction exponents 
in the 4D RSIM. Entries in boldface come directly from the cited references.
The remaining entries come from the scaling relations for logarithmic 
corrections.}
\label{tab:4D}
\end{table}
Using RG, Aharony  derived the unusual exponential terms in (\ref{C1})--(\ref{xiinfty}), and~\cite{Ah76} 
\begin{equation}
 \hat{\alpha}= \frac{1}{2}\,, \quad \hat{\gamma} = 0\,, \quad \hat{\nu}=0\,.
\label{Ah76}
\end{equation}
In~\cite{Boris4D}, Shalaev refined Aharony's calculations to higher order in 
perturbation theory, yielding   
\begin{equation}
\hat{\alpha}= 1.2368\,, \quad \hat{\gamma} = -0.3684\,, \quad \hat{\eta}=0094\,.
\label{Borisalpha}
\end{equation}
Jug's calculations along the $\alpha = 0$ line in $(n,d)$ space 
yielded \cite{GJ83}
\begin{equation}
\hat{\alpha}= 1/2\,, \quad \hat{\gamma} = 1/212 \approx 0.0047\,
\label{Jugalpha}
\end{equation}
at  $(n,d) = (1,4)$. 
In~\cite{GeDe93}, Geldart and De'Bell derived
\begin{equation}
 \hat{\alpha}\approx 1.2463, \quad \hat{\gamma} \approx -0.3684, \quad \hat{\eta}=\frac{1}{212}=0.0047,
\label{GaDe93}
\end{equation}
and finally Ballesteros et al.~\cite{BaFe98} gave
\begin{equation}
 \hat{\alpha}= \frac{1}{2}, \quad \hat{\gamma} = \frac{1}{106} \approx 0.0094, \quad \hat{\nu}=0, \quad \hat{q}=\frac{1}{8}.
\label{BaFe98}
\end{equation}
From these fragmented pictures, complete scaling scenarios may be built
using the scaling relations for logarithmic corrections. 
These scenarios are summarized in Table~\ref{tab:4D}.

It is remarkable that {\emph{none}} of the five analyical works
on this model agree regarding the detail of the logarithmic correction
exponents, a circumstance which motivated our investigations into the
model. We chose to investigate the random-{\emph{site}} versions of 
the models, as this has hitherto proved most controversial in 2D. 
These investigations use the Lee-Yang zeros of the partition function,
which we next describe.

\section{A Fresh Perspective: Lee-Yang Zeros}

We have examined the RSIM in both two and four dimensions from a fresh 
perspective, namely using the Lee-Yang zeros of the partition function.

A phase transition is a physical manifestation of a mathematical 
non-analycity in the free energy. Since the free energy is essentially
the logarithm of the partition function, non-analyticities arise when the 
latter vanishes. Writing the partition function for a  system of size $L$
as $Z_L(t,h)$, one may consider zeros in either of the
variables $t$ or $h$. In each case,
the zeros are located in the complex plane. 
This tack was first suggested by Lee and Yang, 
and complex-$h$ zeros (which, according to a theorem by the 
same authors, are usually located along the imaginary
$h$-axis) are called Lee-Yang zeros. Above the critical temperature 
($t>0$), where there is no phase transition, they are located away from 
the critical point $h=0$. There, the zeros may be considered to be
``proto-critical points'', in the sense that they have the potential to become actual transition
points.

\subsection{Scaling}

The start of the distribution of complex-$h$ zeros
in the $t>0$ phase is called
the Yang-Lee edge,  which we write as $r_{\rm{YL}}(t)$.
As the temperature is reduced towards the critical one ($t \rightarrow 0$),
the edge and the distribution of zeros move towards the real axis,
which they pinch at $t=0$. 
This pinching, which precipitates  the phase transition, scales
in a manner characterised by critical exponents. Allowing for 
logarithmic corrections here too, we write
\begin{equation}
r_{\rm{YL}}(t)  \sim   t^{\Delta}  |\ln{t}|^{\hat{\Delta}} \,,
\label{Delta}
\end{equation}
in the 2D case. For the 4D model, we have to account for the unusual 
correction terms. Following the theory presented in \cite{KeJo06a},
it turns out that the scaling behaviour for the Yang-Lee
edge is (see also \cite{RL})
\begin{equation}
 r_{\rm{YL}}(t) \sim t^{\Delta} \exp{\left({-\frac{3}{2}\sqrt{\frac{6}{53}|\ln{|t|}|}}\right)} |\ln{t}|^{\hat{\Delta}}
\,.
\label{r}
\end{equation}
The exponent $\Delta$ is called the gap exponent. It is related to the
other exponents via 
\begin{equation}
 \Delta = \beta +\gamma\,.
\label{SR5}
\end{equation}
The logarithmic analogue to this scaling relation is \cite{KeJo06a}
\begin{equation}
 \hat{\Delta} = \hat{\beta} - \hat{\gamma}
\,.
\label{SRlog5}
\end{equation}

Besides the scaling behaviour of the Yang-Lee edge, one may also consider
the {\emph{density}} of  zeros, which, for an infinitely large
system, we write as $g_\infty (r,t)$, where $r$ parametrizes their 
locus along the imaginary $h$-axis (assuming the Lee-Yang theorem holds).
In fact it is more convenient to consider the integrated, or cumulative,
distribution function of zeros, which is defined as
\begin{equation}
 G_\infty(r,t) = \int_{r_{\rm{YL}}(t)}^r{g_\infty(s,t)ds}
\,.
\end{equation}
From \cite{KeJo06a,KeJo06b}, in the 2D case at $t=0$ this is 
\begin{equation}
 G_\infty (r) \sim r^{\frac{2-\alpha}{\Delta}} 
(\ln{ r })^{\hat{\alpha}-1-(2-\alpha)\frac{\hat{\Delta}}{\Delta}} 
\,.
\label{G4}
\end{equation}
Following a similar approach to that oulined in \cite{KeJo06a},
in the 4D case, one determines
\begin{equation}
 G_\infty (r) \sim r^{{\frac{2-\alpha}{\Delta}}}
\exp{ \left({
             \left({
                    1 -\frac{3\gamma}{2\Delta} }\right)\sqrt{\frac{6}{53}|\ln{r}}|
      }\right)
    } 
|\ln{ r }|^{\hat{\alpha}-(2-\alpha)\frac{\hat{\Delta}}{\Delta}}
\,.
\label{G5}
\end{equation}
Using the mean-field values $\gamma=1$ and $\Delta = 3/2$, the
exponential term drops out of this expression.

\subsection{Finite-Size Scaling}

The finite-size scaling (FSS) behavior for the susceptibility,
the Yang-Lee edge and the specific heat for each model is determined by 
using (\ref{xi}) and (\ref{xiinfty}) to write the reduced temperature 
in terms of the correlation length. Then the behaviour of the various 
functions may be expressed directly in terms of $\xi_\infty(t)$. 
For sufficiently large lattices,  $\xi_\infty(t)$ may be replaced
by  $\xi_L(0)$ at the critical point. The relationship between 
$\xi_L(0)$ and the lattice size is, in turn, given by (\ref{qhat}),
where ${\hat{q}} = 0$ in the 2D case and  ${\hat{q}}$ is listed
in Table~\ref{tab:4D} in the 4D scenarios.

The unusual exponential correction terms in 4D, 
which otherwise swamp the logarithmic corrections, are not 
in doubt in the literature. To probe the contested logarithmic 
correction terms, therefore, these terms have to be removed.
It turns out that these terms cancel 
out in the FSS expressions for the susceptibility, the Yang-Lee edge 
and the density of zeros in the 4D case.

\begin{figure}
  \includegraphics[height=.2\textheight]{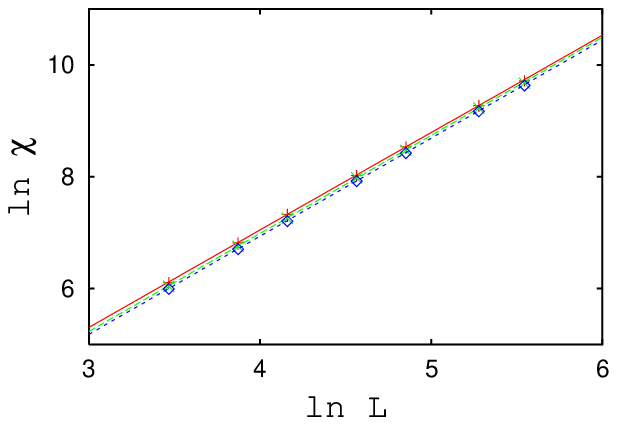}
  \includegraphics[height=.2\textheight]{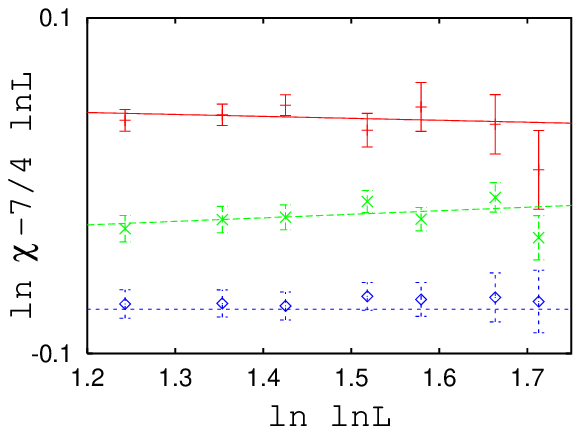}
  \caption{The leading FSS (left) and corrections (right) 
for the susceptibility in the RSIM in two dimensions for weak 
($p=0.88889$, upper data set, red online), moderate ($p=0.75$, 
middle set, green online) and strong ($p=0.66661$, lower set, blue online) dilution values.}
\label{Fig1}
\end{figure}
To summarize, the FSS behaviour for the susceptibility and for the edge 
(i.e., the first Lee-Yang zero)
in both models is given by
\begin{eqnarray}
\chi_L(0)  \sim  L^{\frac{\gamma}{\nu}} 
  |\ln{L}|^{\hat{\zeta}}
&
 \quad \mbox{where} \quad
&
 \hat{\zeta} =
{\frac{\nu {\hat{\gamma}} - \gamma \hat{\nu}+ \gamma \hat{q}}{\nu}}
\,,
\label{zeta}
\\
r_1(L)  \sim   L^{-\frac{\Delta}{\nu}} 
  |\ln{L}|^{\hat{\rho}}
 &
 \quad \mbox{where} \quad 
 &
 \hat{\rho}
 =
 \frac{\nu \hat{\Delta} + \Delta \hat{\nu} - \Delta \hat{q}}{\nu}
\,.
\label{rho}
\end{eqnarray}
The FSS of the specific heat in each model is 
\begin{eqnarray}
 c_L(0) & \sim & (\ln{L})^{\hat{\alpha}} + {\rm{constant}}
 \quad {\mbox{when}} \quad d=2 \,,
 \label{cd=2}
 \\
c_L(0)   & \approx & A           - B^\prime   
\exp{\left({ -2\sqrt{\frac{12}{53} \ln{L}} }\right)}    (\ln{L})^{\hat{\alpha}}
 \quad {\mbox{when}} \quad d=4 \,.
\label{C3}
\end{eqnarray}
In the 2D case, it is difficult to test the Ansatz (\ref{cd=2}), as explained
around Eq.(\ref{controversy}). There, we shall instead extract 
$\hat{\alpha}$ using 
the form (\ref{G4}) for the density, together with the scaling relations 
for logarithmic corrections.

\section{Numerical Approach}

\begin{figure}
  \includegraphics[height=.2\textheight]{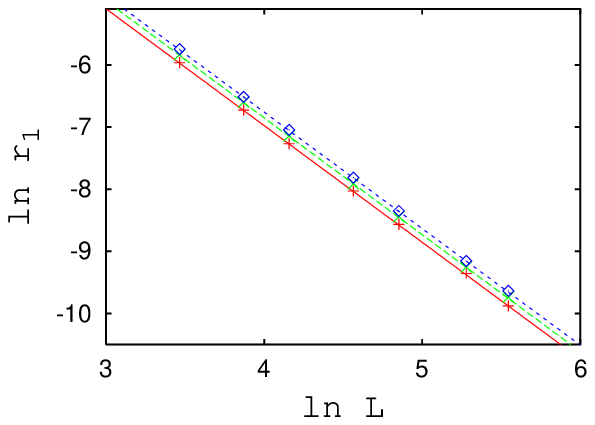}
  \includegraphics[height=.2\textheight]{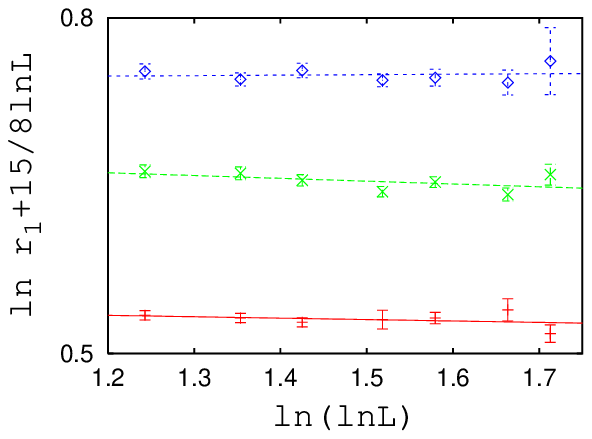}
  \caption{The leading (left) and corrections  (right) to FSS 
 for the Yang-Lee edge for the RSIM in 2D for weak 
($p=0.88889$, lower data set, red online), moderate ($p=0.75$, 
middle set, green online) and strong ($p=0.66661$, upper set, blue online) dilution values.}
\label{Fig2}
\end{figure}
We have simulated the RSIM in both $d=2$ and $d=4$ dimensions using
the Wolff single-cluster algorithm \cite{Wolff89}. 
The quenched dilution is implemented by occupying a site
at a given point on the lattice with probability $p$, so that 
$p=1$ corresponds to the pure (intact) model.
In the $d=2$ case, we simulated at weak, moderate and strong site-dilution
values, represented by $p=0.88889$, $p=0.75$ and $p=0.66661$, respectively,
and we used lattices of size $L=32$, 48, 64, 96, 128, 196 and 256. 
In the $d=4$ case $p=0.8$ and $p=0.5$ with 
$L=$ 8, 12, 16, 24, 32 and 48 were used. We also simulated the 
pure $p=1$ models. The simulations were carried out at the critical 
temperatures estimated in \cite{BaFe97,BaFe98} and periodic boundary 
conditions were used. Up to 1000 realizations of disorder were 
generated for each lattice size and each dilution value, with the 
specific heat, susceptibility, and lowest zeros being determined for
each one. These were then averaged to give estimates for each quenched model.
Further details are supplied in 
\cite{KeRu08,GoKe09}.

\subsection{The Two-Dimensional Case}

We begin the analysis with the susceptibility, the FSS for
which is given in (\ref{zeta}) with the JSSL predictions 
(\ref{leading}) giving  $\gamma/\nu=7/4$.
In fact although the weak hypothesis advocates dilution-dependent 
leading exponents, the ratio $\gamma/\nu$ is believed to be fixed 
there, so one cannot distinguish between the strong and weak
scenarios on this basis. However, Roder et al. have presented 
compelling evidence for the value  $\gamma=7/4$ 
 \cite{RoAd98}, which we can input into FSS in order
to extract $\nu$.
The results of this process are depicted in 
Fig.~\ref{Fig1} and summarized in Table~\ref{tab:r}.
From fits to the leading behaviour $\chi_L(0) \sim L^{\gamma / \nu}$,
the theoretical value  $\gamma/\nu=7/4$ and hence $\nu = 1$ is clearly
supported for each dilution value.
Accepting this value, and fitting for the logarithmic-correction 
exponent $\hat{\zeta}$, one obtains values compatible with zero,
and therefore compatible with theory. 
Then inputting the value $\hat{\gamma}=7/8$ 
clearly evidenced in \cite{RoAd98},
one obtains from (\ref{zeta}) and (\ref{leading}),
values for $  \hat{\nu} - \hat{q} $ clearly compatible with 
the JSSL value $1/2$
(see Table~\ref{tab:r}). The scaling relation (\ref{SRlog1}) then leads
to the estimates $\hat{\alpha}=-0.01 \pm 0.04$, 
$\hat{\alpha}=0.02 \pm 0.03$, and $\hat{\alpha}=0.01 \pm 0.04$, 
for $p=0.88889$, $p=0.75$, and for $p=0.66661$, respectively.

\begin{figure}
  \includegraphics[height=.2\textheight]{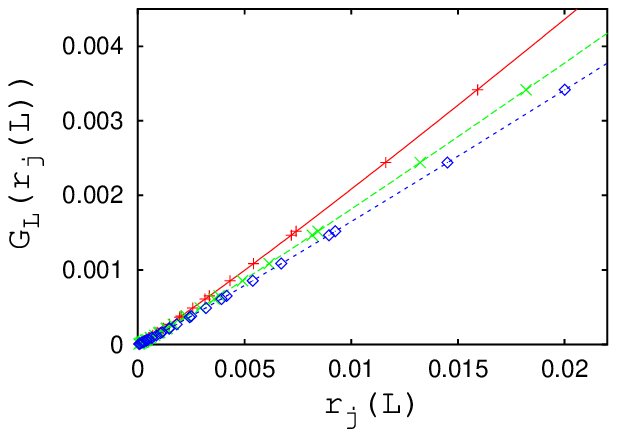}
  \includegraphics[height=.2\textheight]{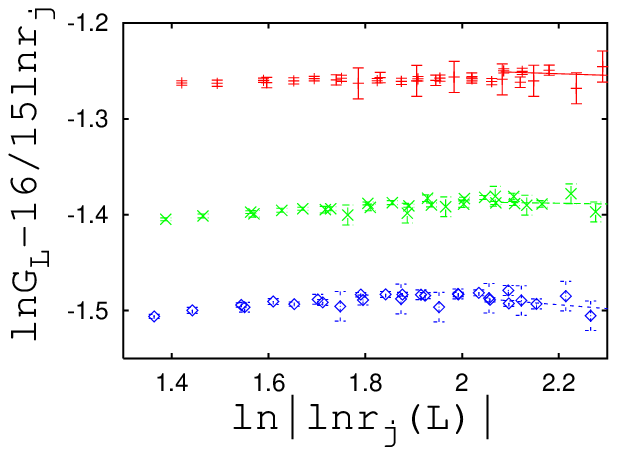}
  \caption{The leading (left) and corrections to (right) finite-size 
scaling for the Yang-Lee edge for the RSIM in two dimensions for weak 
($p=0.88889$, upper data set, red online), moderate ($p=0.75$, 
middle set, green online) and strong ($p=0.66661$, lower set, blue online) dilution values.}
\label{Fig3}
\end{figure}
The FSS for the lowest lying Lee-Yang zeros are plotted in Fig.~\ref{Fig2}
and the results of the corresponding fits are 
also listed in Table~\ref{tab:r}.
The fits to the leading power-law behaviour are consistent with 
the theoretical value  $\Delta /\nu = 16/8$. 
Accepting this leading value, and fitting
for the corrections lead to values of $\hat{\rho}$ compatible with zero.
From Eq.(\ref{rho}), then, one can estimate $\hat{\Delta}$ using the 
established values of $\nu$, $\Delta$ and $\hat{\nu}-\hat{q}$. 
These estimates for $\hat{\Delta}$ are listed in Table~\ref{tab:r}.
\begin{table}[b]
\begin{tabular}{llccc} 
\hline
  \tablehead{1}{l}{b}{Exponent} 
& \tablehead{1}{l}{b}{Theoretical\\value} 
& $p=0.88889 $ 
& $p=0.75$  
& $p=0.66661$ \\
\hline
$\gamma/\nu$  &  $7/4  = 1.75 $  
   & $ 1.747  \pm 0.007$   &   $1.755 \pm 0.005$ &   $1.752 \pm 0.007 $  \\
$\Rightarrow \nu$  &  $1$  
   & $ 1.002  \pm 0.004$   &   $0.997 \pm 0.003$ &   $0.999 \pm 0.004 $  \\
\hline
$\hat{\zeta}=(\nu {\hat{\gamma}} - \gamma \hat{\nu}+ \gamma \hat{q})/\nu$  & 0
   & $-0.01   \pm 0.03$~~~     &    $0.02 \pm 0.03 $ &   $0.01 \pm 0.03$     \\
$\Rightarrow \hat{\nu}-\hat{q}$  & 1/2
   & $0.51   \pm 0.02$     &    $0.49 \pm 0.02 $ &   $0.49 \pm 0.02$     \\
\hline
$\Delta/\nu$  & $15/8 = 1.875$
   & $ 1.879 \pm 0.004$    &   $ 1.878 \pm 0.006  $ &   $ 1.878 \pm 0.006 $  \\ 
$\hat{\rho}=(\nu {\hat{\Delta}} + \Delta \hat{\nu}- \Delta \hat{q})/\nu$  & 0
   & $-0.01   \pm 0.02$     &    $0.04 \pm 0.02 $ &   $0.01 \pm 0.03$     \\
$\hat{\Delta}$  & -15/16 = - 0.9375    & $ -0.95  \pm 0.02$   &   $-0.95 \pm 0.03$ &   $-0.95 \pm 0.03 $  \\
\hline
$\alpha$  & 0                                                   & $ -0.02  \pm 0.03$   &   $0.01 \pm 0.02$ &   $0.00 \pm 0.03 $  \\
$\hat{\alpha}$      & 0                                               & $ -0.02 \pm 0.05$    &   $-0.01 \pm 0.03$ &   $-0.04 \pm 0.05 $  \\ 
\hline
\end{tabular}
\caption{Estimates for critical and logarithmic-correction 
from fits to the scaling behaviour of the susceptibility and first Lee-Yang
zeros for the 2D RSIM. 
These estimates agree with the JSSL theoretical values. }
\label{tab:r}
\end{table}

To summarize the situation at this point, analyses of the FSS of the
susceptibility and the Yang-Lee edge yield results which are 
dilution-independent for both the leading and logarithmic-correction 
exponents and fully compatible with the theoretical predictions of JSSL.

We finally turn to the density of zeros (\ref{G4}). For large enough lattice
size, it is expected that the integrated density $G_\infty(r)$ may be 
estimated from finite lattices as \cite{KeJo06a}
\begin{equation}
 G_L(r_j(L)) = \frac{2j-1}{2L^d} \,,
\label{GL}
\end{equation}
where $j$ is the index of the zero $r_j(L)$.
We note that unlike the specific heat, there is no constant term 
contaminating the expression  (\ref{G4}), allowing  $\alpha$ and 
$\hat{\alpha}$ to be cleanly extracted (having already 
established $\Delta$ and $\hat{\Delta}$).

The density distribution function of zeros is plotted in Fig.~\ref{Fig3}
using the first four zeros for lattices from size 
$L=32$ to $L=256$ ($28$ points in all)
for each value of the dilution. Excellent data collapse
is evident in each case, and
 fits indicate that each curve goes through the origin,
as it should at the critical value of the temperature.
The potential logarithmic corrections are firstly ignored and 
fits to the leading scaling of the density are made. 
For $p=0.88889$, fits to the eight lowest data points
yield and $(2-\alpha)/\Delta =1.076(16)$, 
corresponding to the estimate $\alpha = -0.02(3)$. 
The corresponding results in the $p = 0.75$ and $p=0.66661$ cases are 
 $(2-\alpha)/\Delta =1.062(10)$ ($\alpha = 0.01(2)$) and
 $(2-\alpha)/\Delta =1.066(15)$ ($\alpha = 0.00(3)$), respectively.
These results for $\alpha$ are gathered in Table~\ref{tab:r}.
We next accept the theoretical value $\alpha = 0$ for each 
dilution and explore
potential multiplicative logarithmic corrections by plotting 
$\ln G - 16/15 \ln r$ against $\ln{(\ln{r})}$ in Fig.\ref{Fig3}.
The Ansatz (\ref{Gend}) becomes
\begin{equation}
G_\infty(r) \sim r^{\frac{16}{15}}(\ln{r})^{\hat{\alpha}}\,.
\label{Gend}
\end{equation} 
A fit to all data points 
for $p=0.88889$ gives $\hat{\alpha} =   0.012(3)$, 
four standard deviations from the
theoretical value of zero. However, focusing on the scaling 
region closer to the origin establishes
compatibility with the DDJ value. 
For example, fitting to the lowest eight data points yields 
$\hat{\alpha}  = -0.02 (5)$. 
The equivalent results for $p=0.75$ and $p=0.66661$ are 
$\hat{\alpha} = -0.01(3)$ and $-0.04(5)$, respectively. 
The corresponding fits are depicted in Fig.~\ref{Fig3}
 and the estimates for $\hat{\alpha}$ are summarized in Table~\ref{tab:r}.
These values constitute  numerical evidence that $\alpha= \hat{\alpha}=0$,
independent of dilution and in favour of DDJ and JSSL.

\subsection{The Four-Dimensional Case}

\begin{figure}[t]
  \includegraphics[height=.2\textheight]{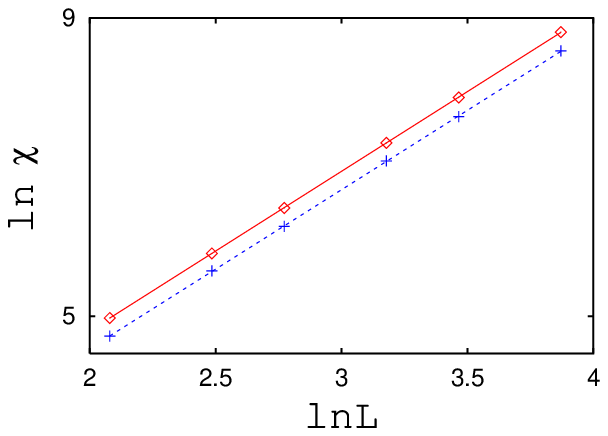}
  \includegraphics[height=.2\textheight]{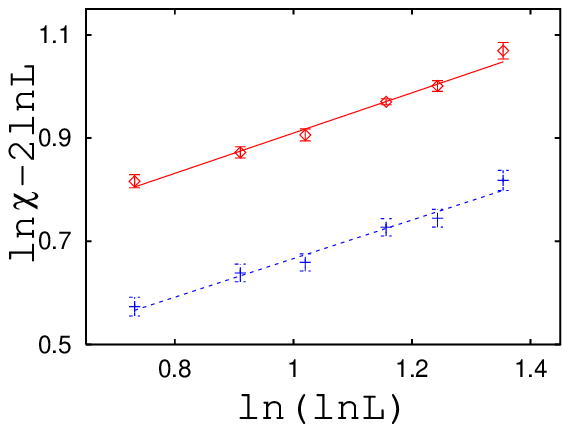}
  \caption{ The leading FSS (left) and corrections (right) 
for the susceptibility in the RSIM in 4D for weak 
($p=0.8$, upper data set, red online) and
 strong ($p=0.5$, lower set, blue online) dilution values.}
\label{Fig4}
\end{figure}

Starting with the weaker dilution value $p=0.8$, ignoring 
logarithmic corrections, and fitting to the leading form of (\ref{zeta})
yields the estimate $\gamma/\nu =  2.14 \pm     0.01$ 
using lattice sizes $L=8-48$ (see Fig.~\ref{Fig4}).
Attributing the deviation from the mean-field value $\gamma/\nu=2$ 
to the logarithmic corrections, we find an appropriate fit 
yields $\hat{\zeta} = 0.39(3)$ for $8 \le L \le 48$.
Thus the FSS logarithmic corrections have moved from the pure value
$\hat{\zeta} = 0.5$ towards the theoretical estimates 
for the diluted value, namely  $\hat{\zeta}\approx 0.25$
to $0.26$.
The same analysis for the FSS of the susceptibility at the stronger
dilution value $p=0.5$ gives similar results: the leading exponent
is estimated at $\gamma/\nu =  2.13 \pm     0.02$ and 
the correction exponent is  $\hat{\zeta} = 0.37(4)$ for $8 \le L \le 48$.
These results are summarised in Table~\ref{tab:s}, 
together with results obtained
from the same fits with the smallest lattices removed. 

\begin{figure}[b]
  \includegraphics[height=.2\textheight]{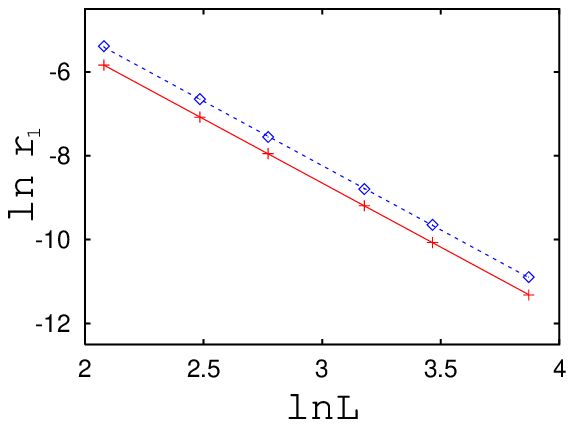}
  \includegraphics[height=.2\textheight]{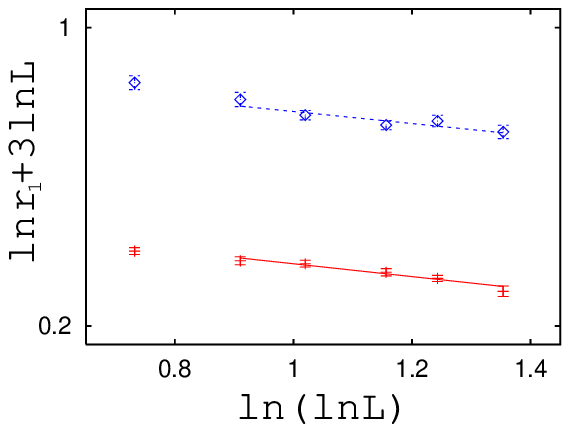}
  \caption{The leading FSS (left) and corrections (right) 
for the Yang-Lee edge in the RSIM in 4D for weak 
($p=0.8$, lower data set, red online) and
 strong ($p=0.5$, upper set, blue online) dilution values.}
\label{Fig5}
\end{figure}

The FSS behaviour for the Yang-Lee edge is plotted in Fig.~\ref{Fig5}.
Fitting only to the leading behaviour in  (\ref{rho}),
for the weaker dilution given by $p=0.8$, one obtains 
$\Delta/\nu = 3.055(8)$,  using all lattice sizes
and at the  stronger dilution value $p=0.5$ we obtained 
$\Delta/\nu = 3.068(13)$.
Again, we interpret these as supportive of the Gaussian
leading behaviour $\gamma/\nu=3$ with logarithmic corrections.

These logarithmic-correction exponents are estimated by fitting to 
(\ref{rho}), with the various theories in the literature indicating that 
$\hat{\rho}= -0.125$ to $-0.13$. We find clean evidence in support of this
with the estimates $\hat{\rho} = -0.15(2)$ and $\hat{\rho} = -0.20(4)$
at $p=0.8$ and $p=0.5$ respectively, using $L=8$--$48$.
Dropping the smallest lattice sizes from the FSS analysis yields 
even more convincing results, namely $\hat{\rho} = -0.17(4)$ 
and $\hat{\rho} = -0.16(5)$ at weak and strong dilution, respectively.
Each of these are supportive of \cite{GJ84,Ah76,Boris4D,GeDe93,BaFe98} 
and   are summarised in Table~\ref{tab:s}.

\begin{figure}
  \includegraphics[height=.2\textheight]{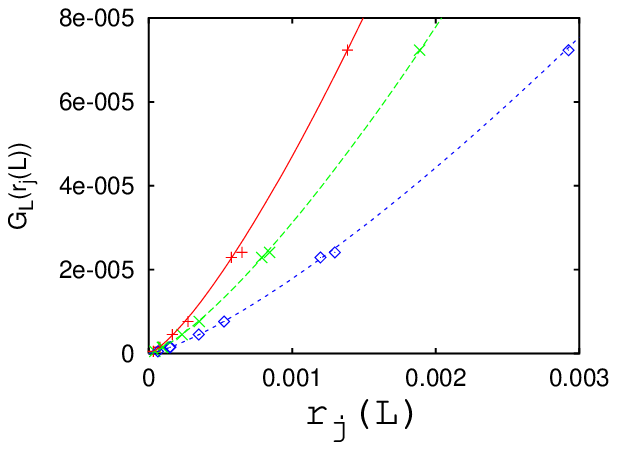}
  \includegraphics[height=.2\textheight]{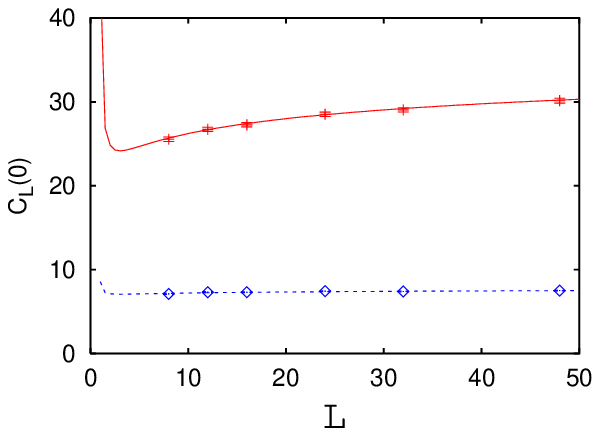}
  \caption{Left: The integrated density of zeros for the pure 
Ising model (upper curve, red online) and the RSIM in 4D with $p=0.8$
(middle curve, green online) and $p=0.5$ (lower curve, blue online).
Right: The specific heat for $p=0.8$ (upper curve, red onlune) 
and for $p=0.5$ (lower curve, blue online) in the 4D RSIM.}
\label{G4D}
\end{figure}

The integrated densities of zeros is calculated in a similar manner 
to the 2D case and are plotted in Fig.~\ref{G4D} for $p=0.8$ and $p=0.5$ 
alongside the equivalent in the pure $p=1$ case. 
A fit to the leading behaviour $G(r) \sim r^{(2-\alpha)/\Delta}$
yields $(2-\alpha)/\Delta = $ 1.32(3), 1.32(1) and 1.32 (1) for
$p=1$, $p=0.8$, and $p=0.5$ respectively.
These are compatible with the theoretical value $4/3$,
independent of dilution strength. 
The errors are too large
for us to be confident about the equivalent density analysis 
for the logarithmic corrections and instead we examine the specific heat 
directly, albeit in a rather unusual manner.

Differentiating Ansatz (\ref{C3}) for the specific heat, one finds
its slope  vanishes when  $C_L(0) = A$ (the asymptote 
$L \rightarrow \infty$) and when 
$L = \exp{ ( (53/12) \hat{\alpha}^2  )}$.
The specific heat is plotted for the two dilution values in Fig.~\ref{G4D},
from which it is clear that the second occurrence of zero slope
is for a 
lattice size smaller than $L=8$.
On this basis, one may conclude
$ \hat{\alpha}
{\rm{\raisebox{-.75ex}{ {\small \shortstack{$<$ \\ $\sim$}} }}}
\sqrt{12/53}\sqrt{\ln{8}}
\approx 0.7
$, and therefore exclude the values $ \hat{\alpha}\approx 1.237$ 
and $\hat{\alpha}\approx 1.246$ given in~\cite{Boris4D,GeDe93}.
The specific-heat curves plotted in Fig.~\ref{G4D}
are best fits to the Ansatz (\ref{C3}) 
with $\hat{\alpha}$ fixed  $1/2$, the alternative
value from \cite{GJ83,Ah76,BaFe98}. 
\begin{table}[b]
\begin{tabular}{l|lll|ll} \hline
$L$-range  &  & $8-48$       & $12-48$  &  $8-48$  & $12-48$  \\ 
\hline
   $p$     &  & \multicolumn{2}{c|}{$\hat{\zeta}$}&\multicolumn{2}{c}{$\hat{\rho}$}  \\ 
\hline
   $0.8$     &  &      $0.39 \pm 0.03$ & $0.42 \pm 0.04$ &$-0.15 \pm 0.02$ &$-0.17 \pm 0.04$ \\ 
\hline
   $0.5$     &  &      $0.37 \pm 0.04$ & $0.40 \pm 0.06$ &$-0.20 \pm 0.04$&$-0.16 \pm 0.05$ \\ 
\hline 
\end{tabular}
\caption{Estimates for FSS exponents for various weak and strong
dilution. One expects that  
$\chi_L \sim L^2 (\ln{L})^{\hat{\zeta}}$ 
where  $\hat{\zeta} \approx 0.25$ to $0.259$
and $r_1 \sim L^{-3} (\ln{L})^{\hat{\rho}}$, 
where $\hat{\rho} \approx -0.125$ to $-0.130$. 
(In contrast, the pure $p=1$ theory is known to have 
$\hat{\zeta} = 1/2$ and $\hat{\rho} = -1/4$.)
}  
\label{tab:s}
\end{table}

\section{Conclusions}

We have presented reviews on the quenched-disordered Ising model in two
and four dimensions, where the specific-heat exponent of the pure
models vanish and no clear Harris prediction 
for critical behaviour at the phase transitions can be made. 
This circumstance
has resulted in both the 2D and 4D models being controversial. 

In the 2D case, the debate has persisted for over thirty years. 
After confirming the JSSL predictions \cite{GJ83,Boris2D,SL,JuSh96} 
for the logarithmic corrections to scaling for the 
susceptibility in the random-site version of that model, 
we determined the  Lee-Yang zeros to high accuracy and 
verified that their 
logarithmic corrections also accord with the JSSL scenario.

In the 2D model, the precise behavior of the specific heat
has been especially controversial and notoriously difficult to pin down directly. 
Using recently  developed scaling relations for logarithmic corrections
\cite{KeJo06a,KeJo06b}, 
together with FSS and Lee-Yang zeros, 
we have presented an alternative approach,
which strongly favours the DDJ  scenario
 \cite{DD,GJ83,GJannouncement,GJ84,Boris2D,SL,JuSh96}.
These analyses were carried out at weak, moderate and strong 
dilution values, thereby supporting of the  strong scaling hypothesis
that the exponents are dilution independent.

In  the 4D case, our analysis  also strengthens
the analytical predictions that
the Gaussian fixed point of the pure model dominates scaling 
and that the logarithmic corrections in the RSIM differ from 
those in the pure model as predicted 
in~\cite{GJ83,Ah76,Boris4D,GeDe93,BaFe98}.
Furthermore, we have succeeded in discriminating between 
some of the detailed analytic 
predictions in the literature, and and our analysis 
favours the predictions 
of~\cite{GJ83,Ah76,BaFe98} over those of~\cite{Boris4D,GeDe93}.


\begin{theacknowledgments}
RK would like to thank the Statistical Physics Group at the 
Institut Jean Lamour, Nancy Universit{\'{e}}, France,
for hospitality during completion of this work as well as
Arnaldo Donoso, Giancarlo Jug and Boris Shalaev for stimulating 
discussions. 
This work has been partly supported by MEC, contracts
FIS2007-60977 and FIS2006-08533-C03. Some of the simulations
were performed in  the BIFI and CETA-CIEMAT clusters.
\end{theacknowledgments}





\end{document}